%% file: paper.tex
\documentclass[sigconf]{acmart}

\renewcommand\footnotetextcopyrightpermission[1]{} 
\usepackage{pifont,csquotes}
\usepackage{enumitem}
\usepackage[official]{eurosym}
\usepackage[multiple]{footmisc}
\usepackage{epsfig,endnotes}
\PassOptionsToPackage{hyphens}{url}
\usepackage{url}
\usepackage{color}
\usepackage{subfigure}
\usepackage{tikz}
\usepackage{listings}

\usepackage{tcolorbox}

\hypersetup{
  colorlinks,
  linkcolor={red!50!black},
  citecolor={blue!50!black},
  urlcolor={blue!50!black}
}
\hypersetup{citecolor=blue,linkcolor=blue}

\usepackage{colortbl}
\usepackage{multirow}
\usepackage{xcolor}
\usepackage{comment}
\usepackage[normalem]{ulem}
\usepackage{xspace}
\usepackage{soul}
\usepackage{textcomp}
\renewcommand{\footnotesize}{\fontsize{9}{10}\selectfont}
\usepackage[T1, OT1]{fontenc}

\usepackage{calc}

\usepackage{enumitem}
\DeclareTextSymbolDefault{\dh}{T1}

\newenvironment{packed_itemize}{
\begin{itemize}[leftmargin=*]
  \setlength{\itemsep}{4pt}
  \setlength{\parskip}{0pt}
  \setlength{\parsep}{0pt}
  \setlength{\topsep}{2pt}
}{\end{itemize}}

\newenvironment{packed_enumerate}{
	\begin{enumerate}[leftmargin=*,label=(\arabic*)]
  \setlength{\itemsep}{2pt}
  \setlength{\parskip}{0pt}
  \setlength{\parsep}{0pt}
  \setlength{\topsep}{2pt}
}{\end{enumerate}}

\newcommand{\deltwo}[1]{}

\newcommand{\del}[1]{}

\definecolor{ultramarine}{RGB}{153,50,204}
\definecolor{forestgreen}{RGB}{19,160,0}

\newcommand{\tmark}{\ding{115}}%
\newcommand{\Triangle}[0]{{\color{ACMOrange}{\tmark}}\xspace}

\newcommand{\eg}{e.g.,\ }

\newcommand{\ie}{i.e.,\ }

\newcommand{\roas}{\textsf{ROAs}\xspace}

\newcommand{\ours}{\textsf{ASINT}\xspace}
\newcommand{\caida}{\textsf{CA2O}\xspace}
\newcommand{\vs}{\textit{vs.}\xspace}
\newcommand{\orgplus}{\textsf{AS2ORG$+$}\xspace}
\newcommand{\assibling}{\textsf{AS-Sibling}\xspace}
\newcommand{\borges}{\textsf{Borges}\xspace}

\makeatletter
\renewcommand{\sectionautorefname}{\S\@gobble}
\renewcommand{\subsectionautorefname}{\S\@gobble}
\renewcommand{\subsubsectionautorefname}{\S\@gobble}
\makeatother

\title{ASINT: Learning AS-to-Organization Mapping from Internet Metadata}

\author{Yongzhe Xu}
\affiliation{%
    \institution{Virginia Tech}
    \country{USA}
}
\author{Weitong Li}
\affiliation{%
    \institution{Virginia Tech}
    \country{USA}
}
\author{Eeshan Umrani}
\affiliation{%
    \institution{Virginia Tech}
    \country{USA}
}

\author{Taejoong Chung }
\affiliation{%
    \institution{Virginia Tech}
    \country{USA}
}
\input{abstract}

\begin{document}
\maketitle

\input{new-introduction}
\input{new-background}
\input{new-design}
\input{new-results}

\input{usecase}

\input{conclusion}

\bibliography{Bibliography/all.xtx}
\bibliographystyle{abbrv --titlecase as-is --short month --no-field address --no-field month --no-field pages}

\clearpage
\newpage
\appendix
\input{ethics}
\input{reproduce}
\input{prompt}
\input{algorithm}
\input{examples}
\end{document}

%% file: abstract.tex
\begin{abstract}
Accurate AS-to-organization mapping underpins Internet measurement and security, yet registries are fragmented, PeeringDB is narrow, and routing views reflect connectivity rather than ownership.
We take a pragmatic step: \ours integrates curated web evidence with retrieval–guided LLM techniques and strict, evidence-cited validation to infer two relations (aliases and directed parent-child) and then revalidates them conservatively.
To keep the dataset sustainable, we operate a public dashboard and API where operators can inspect per-ASN evidence and submit feedback that seeds refreshes.

At scale, \ours maps 112,172 ASNs into 82,840 organization families and, on overlapping AS sets, yields fewer, larger families with 21–24\% more multi-AS groups than prior datasets (\ie CAIDA AS2Org~\cite{CaidaASOrganization}, \orgplus~\cite{arturi-2023-as2org}, \assibling~\cite{chen-2023-sibling}, and Borges~\cite{selmo-2025-borge}).
Quality is high in practice: \ours achieves a precision of 0.9608, a recall of 0.9915 and an accuracy of 0.9752 under manual validation.
Public deployment further drew operator-submitted reports for 595 ASNs across 106 organizations, with only 6 errors (99.0\% observed clustering accuracy), with feedback coming from network operators across all RIR regions.

Better organization context improves downstream analyses: +27.5\% intra-organization RPKI misconfiguration detections, -9.4\% benign hijack alerts, and -5.9\% corrections to cases mislabeled as IP leasing.

We release code, datasets, and the operator platform with APIs;
given persistent ambiguity in organizational names and the continual evolution of corporate structures, an operator-in-the-loop process is essential; the platform records per-ASN feedback with provenance and incorporates it into periodic refreshes and retraining.
The methodology is model-agnostic and stands to improve further as base LLMs advance.
\end{abstract}

%% file: new-introduction.tex
\section{Introduction}
The Internet is composed of tens of thousands of Autonomous Systems (ASes), yet research and operations must reason about real organizations rather than bare AS numbers.
Who owns which AS, which ASes belong to the same corporate family, and which ASes are subsidiaries or brands of a parent firm directly affect routing security, interconnection, and measurement at scale.
Reliable AS-to-organization (AS-to-org) mapping is therefore a prerequisite for tasks that this community cares about: building cleaner entity graphs, reducing false positives in anomaly and hijack detection, constructing fair evaluation sets for learning systems, and performing organization-level topology, risk, and policy studies \eg RPKI (Resource Public Key Infrastructure)~\cite{rfc6810} and BGP investigations~\cite{chung-2019-rpki, grip, cloudflare-radar, luckie-2013-asrelationship}.

Despite decades of effort, accurate AS-to-org mapping remains difficult at Internet scale.
Registry data are fragmented across the five Regional Internet Registries (RIRs), WHOIS entries are inconsistent or stale, and corporate events such as acquisitions, rebrands, and spinouts outpace manual updates, creating many-to-one and one-to-many relationships between names and ASNs.
BGP- or IRR-only views capture routing links rather than corporate control and can mislead ownership inference~\cite{luckie-2013-asrelationship}.
PeeringDB adds operator-maintained context but is narrow: only about 29.0\% of ASNs appear in our snapshot and its free-text fields are uneven~\cite{PeeringDB}.

Prior datasets improve coverage by consolidating WHOIS and PeeringDB signals, or by mining PeeringDB text for sibling clues, but they remain bounded by these sources and typically encode only a flat ``same organization'' label with limited hierarchy. Thus, they often struggle to capture parent-child structure, rebrands, and cross-RIR unifications~\cite{cai-2010-as, arturi-2023-as2org, chen-2023-sibling, selmo-2025-borge}.
A further challenge is validation: there is no complete ground truth and manual audits do not scale, so mappings risk drifting as organizations evolve.

\paragraph{Integrating Web Mining into AS-to-Org Mapping.} 
In this paper, we take a pragmatic step: we integrate curated open web evidence into AS-to-org mapping and apply well-established NLP and LLM techniques under retrieval with strict, evidence-cited validation.
We introduce \ours, which fuses WHOIS and PeeringDB with crawls of company sites, investor pages, wikis, and news.
We then use standard retrieval, named entity recognition, and instruction-following prompts to turn unstructured text into two kinds of organizational links: aliases and directed parent-child ties.
A conservative second pass revalidates candidate links pairwise with precise rules \eg present-tense evidence for control and exact-name directionality, producing organization families that unify cross-RIR identities and capture acquisitions and subsidiaries that mechanical datasets miss~\cite{lewis-2020-retrieval, jing-2022-ner, cai-2010-as, arturi-2023-as2org}.

Because obtaining ground truth is difficult and organizational structures evolve, we operate a public dashboard and APIs where operators can query by ASN or organization, inspect quoted evidence, and submit per-ASN feedback via a lightweight feedback button with optional comments and URLs.
We announced the site on five RIR mailing lists and received responses covering 595 unique ASNs from 106 organizations; only 6 error reports were filed, indicating both accuracy and practical usability for day-to-day operations.
This feedback also supplies fresh crawl seeds for regular refresh and retraining.

Using this pipeline, we map 112,172 ASNs into 82,840 organization families, substantially increasing coverage of cross-RIR ownership, rebrands, and parent-subsidiary structures compared to registry- and PeeringDB-only baselines~\cite{cai-2010-as, arturi-2023-as2org, chen-2023-sibling, selmo-2025-borge, PeeringDB}.
Better organizational context improves downstream studies: we surface 27.5\% more intra-organization RPKI misconfigurations, reduce benign hijack alarms by 9.4\%, and correct 5.9\% of cases previously labeled as IP leasing that are in fact internal reallocations.
We summarize our contributions as follows:
\begin{packed_itemize}
\item \textit{Problem framing with evidence gaps.}
We cast AS-to-organization mapping as Internet-scale entity resolution and explain why mechanical sources alone are insufficient: fragmented and stale WHOIS across RIRs; BGP/IRR that reflect routing rather than ownership; and PeeringDB's narrow, uneven coverage~\cite{luckie-2013-asrelationship, PeeringDB, cai-2010-as, arturi-2023-as2org, chen-2023-sibling}. 

\item \textit{\ours design: a retrieval-first pipeline that turns noisy web text into precise AS-to-organization structure.}
	\ours integrates WHOIS and PeeringDB with curated open-web text, then applies established components (\eg retrieval, NER, and lightweight instruction-following prompts) to infer two relations: aliases and directed parent-child ties~\cite{lewis-2020-retrieval, jing-2022-ner}.
We then formalize a hierarchical model (families $\rightarrow$ alias groups $\rightarrow$ organizations $\rightarrow$ ASNs) and add a strict second-pass validator that requires present-tense evidence and exact-name directionality, yielding an auditable mapping from unstructured sources.

\item \textit{Coverage at scale and unification over baselines.}
On the full run, \ours maps 112,172 ASNs into 82,840 families; on the 111,942-AS intersection with baselines, it forms fewer, larger families and performs consistent unifications of registry- or PeeringDB-centric clusters.
Across comparisons, the median unification factor is about 2.5-2.6 to 1 and multi-AS families increase by 21-24\%, indicating reduced fragmentation while avoiding partnership-driven overmerges.

\item \textit{Operator-in-the-loop validation and measured quality.}
We operate a public dashboard and APIs where operators query by ASN or organization, inspect quoted evidence, and submit per-ASN feedback.
Across Sept 18$\sim$Oct 6, we received responses for 595 ASNs from 106 organizations with 6 error reports (99.0\% observed clustering accuracy).
Manual evaluation further shows that \ours achieves a precision of \(0.9608\), a recall of \(0.9915\) and an accuracy of \(0.9752\)

\item \textit{Impact on measurement and Web security.}
Organization-aware context improves downstream analyses: +27.5\% intra-organization RPKI misconfiguration detections, -9.4\% benign hijack alarms, and -5.9\% corrections to cases previously labeled as IP leasing but actually internal reallocations~\cite{chung-2019-rpki, cloudflare-radar, grip}.

\end{packed_itemize}

Our findings highlight that integrating web-scale open data with retrieval-guided LLM inference addresses networking problems that remain brittle when pipelines rely only on mechanical sources (\eg WHOIS and PeeringDB).
To support reuse, auditing, and continuous improvement, we make our code, datasets\footnote{https://anonymous.4open.science/r/asint}, and the operator-feedback platform publicly available, together with an API, at:
\begin{center}
    https://asint.netsecurelab.org
\end{center}

%% file: new-background.tex
\section{Background}

\subsection{ASNs, Organizations, and Why the Mapping Matters}
An Autonomous System Number (ASN) is a globally unique identifier assigned by a Regional Internet Registry (RIR) and used by the Border Gateway Protocol (BGP)~\cite{rfc4271} to advertise reachability for IP prefixes.
An ASN represents a routing domain that applies a consistent policy at its borders, but it does not, by itself, reveal legal ownership or corporate control.
Real organizations often operate multiple ASNs across regions, time zones, and business units; conversely, the same brand may appear in registries under different local names after mergers, rebrands, or spinouts.
These many-to-one and one-to-many patterns make plain why mapping ASNs to organizations is critical for security, measurement, and modeling.
For example, RPKI deployment studies depend on correctly attributing invalid announcements to the right owner, hijack detectors must suppress alarms for internal reannouncements, and graph-centric analyses benefit from deduplicating entities before computing organization-level cones and rankings~\cite{chung-2019-rpki, grip, cloudflare-radar, luckie-2013-asrelationship}.
\paragraph{What data exist today and where they fall short.}
Researchers and operators primarily draw on WHOIS records and PeeringDB.
WHOIS provides registry-maintained organization metadata, but fields, naming conventions, and update practices vary by RIR and over time, leading to stale or fragmented identities~\cite{cai-2010-as}.
PeeringDB contributes operator-maintained entries with names, sites, and limited free text, yet coverage is narrow and uneven: in our snapshot only 29.0\% of ASNs appear at all, and many fields are sparsely populated~\cite{PeeringDB}. 

Datasets that consolidate these sources have moved the field forward but remain bounded by them.
CAIDA's AS2Org (\caida) and its extensions group ASNs using WHOIS signals and selected PeeringDB attributes~\cite{cai-2010-as, arturi-2023-as2org}.
\assibling uses PeeringDB and auxiliary signals to infer sibling AS ownership, improving recall in some regions but still constrained by registry and PeeringDB footprints~\cite{chen-2023-sibling}.
Recently, \borges applied LLMs to parse free text and icons from PeeringDB, which increased coverage but kept it tied to that single source~\cite{selmo-2025-borge}.

Most importantly, all of these resources largely encode {\em flat same organization links} and struggle to capture hierarchy such as parent-subsidiary ties, brand aliases, and cross-RIR unifications that matter for downstream tasks.

\paragraph{A web signal that the registry does not capture.}
We note that many ownership facts first appear on the web such as corporate sites, investor pages, and news stories publish acquisitions, rebrands, and subsidiary disclosures long before registries converge.
These texts provide high-value clues missing from WHOIS and PeeringDB, but they are unstructured, noisy, and distributed, which motivates modest NLP to turn them into reliable evidence.
This paper takes that path: bring curated web snippets into the evidence set to complement registries, then use established NLP and instruction-following to turn text into alias and parent-child edges.

\subsection{NLP and Multi-source Inference Essentials for AS-to-ORG}
AS-to-org mapping is an Internet-scale entity-resolution problem over heterogeneous sources.
Mechanical datasets such as WHOIS and PeeringDB offer high-precision anchors but are narrow, inconsistent, or stale; open web text is richer but noisy.
Our design combines these signals with retrieval-guided NLP so that inference is grounded in short, auditable snippets rather than open-ended generation.
We outline the minimal ingredients. 
\paragraph{Retrieval first, generation second.} Retrieval-augmented generation (RAG)~\cite{lewis-2020-retrieval, shuster-2021-retrieval, gao-2023-retrieval} conditions an LLM on small, targeted snippets retrieved for the entities of interest, reducing hallucinations~\cite{huang-2025-hallucination, kaddour-2023-challenges} and keeping the model focused on current facts~\cite{lewis-2020-retrieval}. RAG has shown success in multiple applications~\cite{xu-2024-rag, li-2023-rag, husain-2019-codesearchnet}.
In our setting, retrieval means indexing vetted passages from company sites, wikis, and news, and fetching only those that co-mention candidate organizations with an ASN-linked entity.
\paragraph{Entity discovery and normalization.} Named Entity Recognition (NER) finds candidate organization names in web text, which we then normalize against a canonical list derived from WHOIS and PeeringDB to handle variants \eg legal vs brand names~\cite{jing-2022-ner}. %
Lightweight semantic matching helps merge near-duplicates without overfitting to string equality.
\paragraph{Relation typing with curated prompts.} Few-shot, instruction-style prompts~\cite{zhao-2021-calibrate} ask for one of three relation labels between organization pairs (alias, parent–child, or none) and can improve model accuracy and consistency~\cite{chen-2024-unleashing, sahoo-2024-systematic, kaddour-2023-challenges}.
We rely on evidence sentences retrieved for the specific pair and apply conservative rules: we default to ``none'' without explicit, present-tense mentions, and require exact name matches for directionality when labeling parent-child.
\paragraph{Why this combination is enough.}
Together, these components convert noisy text into structured edges and integrate naturally with mechanical sources: registries ground coverage, web text adds missing relationships, and prompts resolve local ambiguity with human-readable justification.
This compact toolbox is sufficient to discover cross-RIR aliases, acquisitions, and subsidiaries that registries and PeeringDB miss while keeping costs and complexity in check~\cite{cai-2010-as, arturi-2023-as2org, chen-2023-sibling, selmo-2025-borge}.

%% file: new-design.tex
\begin{figure*}[t]
    \centering
    \epsfig{figure=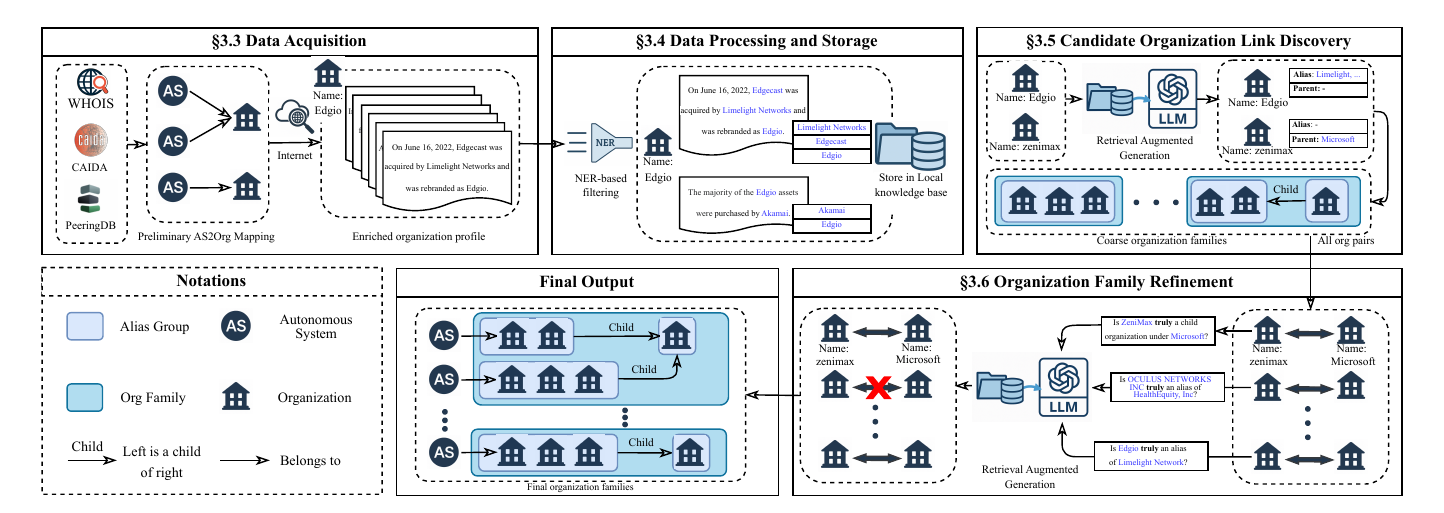,width=1\textwidth}
	\caption{An overview of the \ours\ framework, from data collection to final clustering of organization families.}
    \label{fig:architecture}
\end{figure*}
\section{ASINT: Design}
As shown in \autoref{fig:architecture}, \ours is a conservative, evidence-driven system for mapping ASNs to their real organizations by fusing registry data with curated web evidence and applying well-studied NLP and LLM techniques.
We do not introduce a new algorithm, but the work is not trivial: sources are heterogeneous, records are stale or inconsistent across RIRs, names vary by region and time, and validation must meet operator needs at Internet scale.
Our design goal is to ground every inference in explicit artifacts, keep prompts narrow and auditable, and maintain a feedback loop so operators can correct errors and help future refreshes.
\subsection{Motivation and Scope}
\label{sec:design-motivation}
The design starts from three persistent gaps in AS-to-org mapping.
First, coverage is restricted when relying on WHOIS and PeeringDB alone; organizations change faster than registries update, and only a minority of ASNs appear in PeeringDB~\cite{PeeringDB, cai-2010-as, arturi-2023-as2org, chen-2023-sibling, selmo-2025-borge}.
Second, relationships are oversimplified; most resources mark only ``same organization'' and do not represent aliasing or parent-child hierarchies that matter for measurement and security.
Third, static or rule-based parsing of free text misfires on unstructured descriptions, regional variants, and rebrands.

To address these gaps, \ours augments registry baselines with evidence from company websites, investor pages, wikis, and news, which often publish acquisitions and rebrands before registries converge.
Instead of asking an LLM to ``know'' the world, we retrieve small, targeted snippets and prompt for constrained decisions, keeping the model grounded and auditable~\cite{lewis-2020-retrieval, jing-2022-ner}.
The approach raises challenges that drive the design.
\begin{packed_enumerate}
	\item \textit{Evidence curation at scale}: the web is vast and noisy; we must surface only passages that co-mention candidate entities and the ASN-linked organization.
	\item  \textit{Name normalization across RIRs and languages}: the same operator appears under legal, brand, and regional variants, requiring careful alias handling.
	\item \textit{Pairwise explosion}: naive pairwise inference over tens of thousands of orgs is infeasible; we need coarse pre-grouping to prune the search space, then strict validation to remove false merges.
	\item \textit{Hallucination control and directionality}: LLM outputs must be tied to retrieved text, prefer ``none'' without explicit present-tense evidence, and respect direction for parent-child.
	\item \textit{Operator-grade validation and upkeep}: mappings must be explainable and correctable, with a feedback loop that supports periodic retraining and dataset refresh.
\end{packed_enumerate}
\subsection{Output data model}
\ours produces a hierarchical mapping that supports both flat and structured use cases.
\begin{packed_itemize}
	\item{ \textit{Organization family}}: a top-level unit that contains one or more alias groups under common ownership and optional parent-child links.
	\item{ \textit{Alias group}}: a set of normalized organization records that refer to the same real-world entity under legal, brand, or regional variants.
	\item{ \textit{Organization}}: a normalized record derived from registries or web sources, with stable identifiers, names, websites, and provenance.
	\item{ \textit{ASN}}: a leaf node linked to its owning or operating organization.
\end{packed_itemize}
This model lets downstream tasks reason at multiple levels: flat ``same organization'' for de-duplication, or hierarchical parent-child when ownership structure matters.
\subsection{Data Acquisition}
\label{sec:data-acquisition}
\begin{packed_enumerate}
	\item{\textit{Building the ASN inventory from registries only:}}
We enumerate all public ASNs from bulk WHOIS across the five RIRs and selected NIRs, then reconcile overlaps with as-block authority and RIR precedence.
To close gaps where OrgIDs or linkable handles are missing or inconsistent, we incorporate CAIDA AS2Org as an auxiliary registry view for listing ASNs and seeding initial org records~\cite{cai-2010-as}.
At the end of this step, every ASN has a preliminary organization record derived solely from WHOIS or CAIDA, without using PeeringDB or web data.
\item{\textit{Attaching metadata that can later cue grouping and validation:}}
Once the ASN inventory is fixed, we enrich each ASN and its preliminary organization with optional metadata that helps discover aliases and hierarchies but does not decide them yet.
We use two channels: PeeringDB metadata and targeted web crawl.
For ASNs present in PeeringDB, we attach network name, alternate names, website URL, and operator notes when available, while acknowledging limited coverage and uneven free-text fields~\cite{PeeringDB, arturi-2023-as2org}. 
We also, for every preliminary organization name, issue a small number of focused queries \eg ``parent company'', ``acquired by'', ``subsidiary'', ``rebrand'', and collect high-signal pages from corporate sites, investor pages, wikis, and news.
We extract text, split it into coherent snippets, and store each snippet with URL provenance.
These snippets will be filtered and retrieved later, not treated as ground truth at this stage.
\end{packed_enumerate}

Even with this staging, three problems remain hard.
First, naming is messy: legal and brand variants differ across RIRs and languages, so alias discovery requires careful normalization.
Second, evidence is noisy: free text must be filtered and retrieved in small, relevant snippets to avoid LLM drift.
Third, scale is real: naive pairwise checks across all organizations are intractable, so we need coarse pre-grouping before strict revalidation.
These constraints motivate the subsequent stages: building a compact knowledge base from the enriched snippets, using retrieval-augmented prompts for alias and parent-child typing~\cite{lewis-2020-retrieval}, and validating with conservative rules and operator feedback.

\subsection{Data Processing and Storage}
\label{sec:data-processing}
\ours transforms noisy web pages and registry notes into auditable evidence about organizational ties.
Two principles guide this stage: keep only text that can plausibly support a claim about an organization pair, and preserve provenance so every decision is traceable.
The output is a local knowledge base that indexes short, relevant snippets per organization and per candidate counterparty, enabling precise retrieval in later stages~\cite{lewis-2020-retrieval}.

\paragraph{Notation and objective.}
Let \(\mathcal{A}\) be the set of ASNs and \(\mathcal{O}\) the set of organization records after Step~A in \S\ref{sec:data-acquisition}.
Let \( f_{0} : \mathcal{A} \rightarrow \mathcal{O} \) be the initial registry-based map from ASNs to organizations.
We build a per-organization evidence base \(\mathcal{K}\) and a candidate set \( C(o) \subseteq \mathcal{O} \) for each \( o \in \mathcal{O} \), where every \( c \in C(o) \) is a ``potentially related'' organization that co-occurs with \(o\) in vetted text.
This stage strives for high precision in evidence admission so that later inference can be narrow, grounded, and cheap.

\paragraph{Global name inventory.}
We compile a master list of canonical organization names from WHOIS, CAIDA AS2Org, and PeeringDB, plus straightforward variants \eg legal and brand forms~\cite{cai-2010-as, arturi-2023-as2org, PeeringDB}.
Each canonical name \(n\) is embedded with a sentence transformer \(\phi(n) \in \mathbb{R}^d\) using \texttt{all-MiniLM-L6-v2} to enable fast semantic lookups during filtering~\cite{all-MiniLM}.

\paragraph{Evidence curation pipeline.}
For each target organization \(o \in \mathcal{O}\), we process its collected context through four filters:%
\begin{packed_enumerate}
\item \textit{Text normalization and splitting:}
We strip HTML to text, deduplicate by \(\{\text{URL}, \text{checksum}\}\), then split into short, semantically coherent chunks with small overlap so key phrases are not cut at boundaries.
\item \textit{Named entity recognition and blocking:}
We run NER to extract candidate organization mentions from each chunk, embed those mentions, and keep only candidates whose embeddings are within a conservative FAISS KNN threshold of the global inventory~\cite{jing-2022-ner, faiss}.
\item \textit{Co-mention relevance gating:}
A chunk is retained for \(o\) only if it co-mentions \(o\) and at least one validated candidate \(c\).
Formally, define a binary gate \( G(o,c,s) = 1 \) iff snippet \(s\) contains exact or normalized matches of both \(o\) and \(c\); we keep \(s\) only when \(\exists c : G(o,c,s) = 1\).
\item \textit{Knowledge base construction:}
For every retained snippet \(s\), we store \((o,c,s,\text{URL},t)\) and its embedding \(\phi(s)\) in a vector index keyed by \(\{o,c\}\).
\end{packed_enumerate}

At the end of this stage, each organization \(o\) is linked to a compact set \(\mathcal{K}(o) = \{s_1, s_2, \dots, s_m\}\) of high-signal snippets, from which mentions of other organizations {\em other than} \(o\) are extracted to form a deduplicated candidate set \(C(o) = \{c_1, c_2, \dots, c_n\}\).

\subsection{Candidate Organization Link Discovery}
\label{sec:candidate-links}
Given \(C(o)\) and \(\mathcal{K}(o)\), \ours proposes typed links that later become alias groups and parent-child edges.
We intentionally use simple, constrained prompts over retrieved text rather than open-ended generation to reduce hallucinations, keep direction explicit, and make results explainable.

\begin{packed_enumerate}
\item \textit{Retrieval before inference:}
For each \((o,c)\) with \(c\in C(o)\), we retrieve the top-\(k\) snippets:  
\[
S_{o,c} = \mathrm{TopK}\Big(\{\,s \in \mathcal{K}(o) : G(o,c,s)=1\,\},\; \mathrm{sim}^{*}(s,o,c)\Big),
\]  
where \(\mathrm{sim}^{*}(s,o,c)\) is defined by comparing \(\phi(s)\) with embedded queries such as ``\(o\) is an alias of \(c\)'', ``\(o\) is acquired by \(c\)'', or ``\(o\) is the parent of \(c\)''.
We skip the pair if \(S_{o,c}=\varnothing\).

\item{ \textit{Constrained relation typing:}}
We apply an LLM-based labeling function \( h(o,c,S_{o,c}) \) that returns a label
\( y \in \mathcal{Y} = \{ \texttt{alias}, \texttt{parent}, \\ \texttt{child}, \texttt{none} \} \)
together with a confidence score \(p\) and a cited sentence \(e \in S_{o,c}\), using the prompt described in \autoref{sec:prompt}.

\item{ \textit{Edge scoring and justification:}}
From all labeled pairs we construct a typed, evidence-backed candidate graph
\[
G^{0} = (V, E^{0}_{\mathrm{a}}, E^{0}_{\mathrm{p}}),
\]
where \(V = \mathcal{O}\),
\(E^{0}_{\mathrm{a}} = \{ (o,c,p,e) : h(o,c) = \texttt{alias} \}\) is undirected with weights \(p\), and
\(E^{0}_{\mathrm{p}} = \{ (o \rightarrow c, p, e) : h(o,c) = \texttt{parent} \lor h(o,c) = \texttt{child} \}\) is directed with weights \(p\).

\item{ \textit{Coarse grouping for tractability:}}
We form preliminary alias groups by connecting organizations whose alias edges exceed a conservative threshold and taking connected components in \((V,E^{0}_{\mathrm{a}})\).
We then propose directed parent-child links between alias groups when a consistent majority of member organizations independently point to the same parent.
\end{packed_enumerate}

\subsection{Organization Family Refinement}
\label{sec:org_family_refinement}
Preliminary groups capture many true ties but may admit false merges (as shown in \autoref{sec:false_cluster}).
We therefore rebuild the graph under stricter rules that prioritize precision and interpretability.

\begin{packed_enumerate}
	\item{ \textit{Pairwise revalidation:}}
Within each preliminary alias group \(G\), we enumerate all unordered pairs \((o_i,o_j)\) and re-test them with a stricter pair-wise prompt (shown in \autoref{sec:prompt}); accept \texttt{alias} only with present-tense cues and exact names.
For cross-group control, we re-test every proposed parent-child pair \((G_c \rightarrow G_p)\) by checking all cross-pairs \((o \in G_c, o' \in G_p)\) with LLM.

\item{ \textit{Graph reconstruction:}}
Define the validated alias relation \(\sim\) as the set of pairs judged \texttt{alias} under strict criteria.
Alias groups become cliques of \(\sim\): every pair inside a group must satisfy \(o_i \sim o_j\).
A directed edge \(G_c \rightarrow G_p\) persists only if most cross-pairs validate with the same direction.

\item{ \textit{Invariants and failure handling:}}
We enforce four invariants:
(1) \(\sim\) is an equivalence relation;
(2) parent-child edges are directional and acyclic at the alias-group level;
(3) no node can be both an alias and a strict parent of another;
(4) every surviving edge carries a cited sentence.
Where conflicts arise, we remove the weaker edge or split the group.
\item{ \textit{Final artifact:}}
The refined output is a set of organization families: cliques of aliases plus directed parent-child edges, each with per-edge quotes and URLs.
This design yields an evidence-driven graph that both researchers and operators can trust~\cite{cai-2010-as, arturi-2023-as2org, chen-2023-sibling, PeeringDB}.

\end{packed_enumerate}

%% file: new-results.tex
\section{Results and Analysis}
\input{tables/new-comparison-combined}

\subsection{Datasets and coverage}
Starting from bulk WHOIS across the five RIRs and selected NIRs, augmented with CAIDA's AS2Org for regions lacking explicit OrgIDs, we enumerate 112,172 ASNs and 88,700 preliminary organizations for downstream processing~\cite{cai-2010-as, arturi-2023-as2org}.
PeeringDB contributes operator-maintained metadata but covers only 32,554 ASNs (29.0\%) in our snapshot, underscoring the risk of under-coverage when relying on it alone~\cite{PeeringDB}.
\vspace{-0.05in}
\paragraph{Web evidence footprint.}
To capture organizational relationships that registries miss, we perform targeted web queries per organization name (\eg ``parent company'', ``acquired by'', ``rebrand'') across corporate sites, investor pages, wikis, and news. 
From 873,949 URLs, we extract 20,478,907 raw text chunks, then apply NER~\cite{bert-base-NER} and name normalization to retain 386,278 vetted snippets that co-mention a target and at least one candidate organization~\cite{jing-2022-ner}. %
We then build a local vector database using Milvus~\cite{Milvus} from these filtered chunks, forming a knowledge base that supports subsequent retrieval-augmented generation (RAG) queries~\cite{lewis-2020-retrieval}.

\vspace{-0.05in}
\paragraph{Candidate Org Links Discovery}
In this stage, we use DeepSeek-R1-Distill-Qwen-32B as the LLM model. 
Before inference, each organization has on average 6.97 \textit{candidate organization names}. 
After inference, 2.92 candidates per organization (41.9\%) are classified as either alias or child/parent. 
Overall, this process produces 74,848 organization families.

\vspace{-0.05in}
\paragraph{Org Family Refinement}
By parsing the organization families generated in the previous stage, we construct all organization pairs for pairwise LLM validation, as described in (\autoref{sec:org_family_refinement}). 
In total, we obtain 51,493 alias pairs and 31,907 child/parent pairs for validation, a number far smaller than the total possible organization pairs in the entire search space.
After validation, 9,945 alias pairs (19.3\%) and 3,883 child/parent pairs (12.2\%) remain, highlighting the inaccuracy of purely name-based clustering. 

Finally, we arrive at 82,840 final \textit{organization families}, each comprising AS Numbers under common ownership and control. These clusters provide a comprehensive AS-to-organization mapping that captures rebranding, acquisitions, and shared operational oversight.

\subsection{Evaluation}
\label{sec:evaluation}
Assessing AS-to-organization quality is challenging because no complete ground truth exists and corporate structure evolves.
We therefore evaluate at two complementary levels that reflect how the mapping is used in practice.
\textbf{Level 1: ASN and family membership.} We ask whether each ASN placed in a family truly belongs there.
This lens is well suited to estimating false positives via manual verification and operator feedback, but it cannot directly expose false negatives: an ASN that is missing from a family leaves no trace to audit at the ASN level.
\textbf{Level 2: organization-pair links.} We evaluate pairwise links between organizations (\texttt{alias} or \texttt{parent--child}) that underpin families.
At this level we can estimate both false positives and false negatives by sampling pairs and judging the presence or absence of a link.

\subsubsection{Level 1: ASN and family membership}
\paragraph{Manual verification.}
We randomly sampled 150 families comprising 3,152 ASNs and manually verified membership against the evidence surfaced by \ours.
Under our definition, the resulting family-level false-positive rate is 1.3\%, indicating low overmerge in practice.

\paragraph{Network-operator feedback.}
To measure and reduce errors at scale, we operate a public dashboard and APIs where operators query by ASN or organization, inspect quoted evidence for each decision, and submit lightweight per-ASN feedback with optional comments and URLs.
During the observation window (Sept. 18th to Oct. 6th), the site received 2,940 visits spanning 595 queried ASNs across 106 families; operators filed 6 incorrect-cluster reports, consistent with an observed clustering accuracy of about 99.0\%.
No ``missing link'' reports were filed in the same window; while absence of reports is not proof of completeness, it suggests that many commonly visible relations are already captured by \ours.

\paragraph{Representative error pattern.}
Manual investigation of the operator reports highlights a recurring and intrinsically hard source of error: \textit{extreme name similarity across time or regions}.
A canonical example is the Raiffeisen case: ``Raiffeisen Rechenzentrum GmbH'' versus ``Raiffeisen Informatik GmbH \& Co KG''.
One entity's former name overlaps the other's current legal name, which makes aliasing tempting even for experts when evidence is sparse or historical; the mistake can propagate to subsidiaries, amplifying the error.
A second example is cross-jurisdiction brand reuse: multiple independent firms operate under the moniker ``COMNET'' (\eg PT Comtronics Systems, IPLUS LLC, and Figueiredo Provedores Eireli).
Shared brand strings without present-tense ownership evidence do not imply aliasing or control.
These cases motivate the disambiguation rules we enforce elsewhere: prioritize exact legal-entity names, require co-mention of both legal names in the same sentence for aliasing, penalize cross-country aliases absent explicit evidence, and treat ``formerly known as'' as time-bounded and insufficient on its own.

\subsubsection{Level 2: organization-pair links}
Because pairwise links are the atoms from which families are built, we estimate both false positives and false negatives at the organization-pair level.
We use two complementary probes plus operator signals.

\paragraph{Corpus-level candidate-pair sampling.}
To estimate pair-level precision and bound recall without a gold standard, we proceed:
\begin{packed_itemize}
	\item We first select 150 organization families and enumerate all organization pairs that \ours asserts as linked within those families.
Manual review confirms 1,054 true links and 43 spurious links, yielding a precision of \(0.9608\).

\item We then construct a candidate {\em negative pool} consisting of all organization pairs that share at least one NER-extracted keyword in the filtered web corpus yet do not appear together in the final families; in this way, it targets \emph{findable} misses: if two organizations leave any footprint in public text such as brand, legal, or subsidiary cues, our method ought to have a chance to link them; evaluating recall \emph{conditional on web visibility} is therefore more meaningful than uniform sampling over the combinatorial space of all pairs.
This pool contains 25,068,073 unique pairs; from it, we draw a stratified sample of 1,000 pairs (mixing high-retrieval-score and uniform draws) and manually judge whether each sampled pair is truly related \eg \texttt{alias} or \texttt{parent--child} based on retrieved snippets and linked sources.
We observe 9 missed links in 1,000 samples, implying an estimated false-negative rate of about \(0.9\%\) over this pool and \(\mathrm{TN}=991\).

\item Taken together, our evaluation achieves: \(\mathrm{TP}=1,054\), \(\mathrm{FP}=43\), \(\mathrm{FN}=9\), \(\mathrm{TN}=991\), corresponding to precision \(0.9608\), recall \(0.9915\), specificity \(0.9584\), NPV \(0.9910\), and accuracy \(0.9752\); detailed calculations are provided in \autoref{sec:calculation}.
\end{packed_itemize}

\paragraph{Operator signals.}
The dashboard allows users to flag missing links directly in the family view and attach URLs; across the initial window we received no such reports.
We treat this as supportive but not conclusive and rely on the sampled probes above to quantify residual false negatives.
\subsubsection{Takeaways}
First, family-level precision is high: both manual verification and operator feedback point to low overmerge, with residual errors dominated by extreme name proximity and brand reuse.
Second, when evaluated at the organization-pair level, the mapping achieves high recall and precision on sampled cases, and validator-induced misses are rare and explainable.
Third, the operator-in-the-loop platform is essential for sustainability: it supplies targeted corrections and fresh crawl seeds, and it ensures that the mapping remains auditable and updatable as organizations rebrand, merge, or split.

\subsection{Comparison with Prior Datasets}
\label{sec:comparison-other-datasets}
We benchmark \ours against four widely used resources;
our goal is to show how adding web evidence with conservative, evidence-cited inference reduces fragmentation without introducing partnership-driven overmerges.
The baselines are: (1) \textit{\caida} (a WHOIS-centered mapping~\cite{CaidaASOrganization}); (2) \textit{\orgplus} (\caida enriched with PeeringDB and \texttt{as\_relationship}~\cite{arturi-2023-as2org}); (3) \textit{\assibling} (PeeringDB-centric siblings with support from \caida and BGP~\cite{chen-2023-sibling}); and (4) \textit{\borges} (PeeringDB plus branding-domain merges via an LLM~\cite{selmo-2025-borge}).

We compare \ours to each baseline on overlapping AS sets and report: (1) the number of families and multi-AS families; (2) average family size; (3) \textit{unifications}, where one dataset fully \textit{subsumes} multiple families from the other; and (4) \textit{residuals}, which neither match nor are subsumed.
In the unification rows, the entry ``A / B'' reads as: \textit{A families in the reporting dataset that subsume B families from the counterpart}.
We also include manual validation counts and the true-positive ratio (TPR).

\subsubsection{Overall patterns}
\label{sec:comparison-overall}
As shown in \autoref{tab:compare_all}, across all baselines, \ours consistently forms \textit{fewer, larger} families on the same AS sets, indicating that it unifies entities that registry- and PeeringDB-centric pipelines often leave split.
Against \caida and \orgplus, the mean family size rises from 1.25 to 1.36 (+8.8\%), and multi-AS families rise by roughly 24\% (10,358 vs 8,357 and 10,358 vs 8,327).
Against \assibling and \borges, the mean family size rises from 1.24 to 1.33 and from 1.28 to 1.36, with multi-AS families up by 21-22\%.
These shifts align with our design: web evidence surfaces aliases, subsidiaries, and rebrands that are weakly expressed in registry or PeeringDB text, while strict, evidence-cited validation curbs overmerges.
\subsubsection{Per-baseline findings} \label{sec:per-baseline}
Having seen the overall trends, we now examine each baseline in more detail.
\begin{packed_itemize}

\item{\vs \caida:} On 111,942 shared ASNs, \ours yields 82,631 families vs 89,486 for \caida, with 10,358 vs 8,357 multi-AS families and average size 1.36 vs 1.25.
\ours performs 4,272 unifications that subsume 11,127 \caida families \ie about 2.6-to-1, and there are no reverse unifications or residuals on either side.

\item{\vs \orgplus:}
On the same 111,942 ASNs, \ours forms 82,631 families vs 89,280 with average size 1.36 vs 1.25 and 10,358 vs 8,327 multi-AS families.
\ours unifies 10,929 \orgplus families into 4,232; reverse unifications are 73 that subsume 129 of \ours.
Reverse cases generally reflect features outside our evidence set such as partnership-driven cues from \texttt{as\_relationship}, which are valuable for peering but orthogonal to ownership.
Residuals are small and appear only on \orgplus (8).

\item{\vs \assibling:}
With 108,902 ASNs in common, \ours forms 82,064 families vs 88,129, with 10,272 vs 8,521 multi-AS families and average size 1.33 vs 1.24.
\ours unifies 9,824 \assibling families into 3,732; reverse unifications are 102 subsuming 146 of \ours, with 1 residual on \ours and 20 on \assibling, largely due to snapshot and PeeringDB footprint differences.
Despite \assibling's narrower scope, \ours achieves a higher TPR in manual checks (98.9\% vs 91.9\%).

\item{\vs \borges:}
On 111,937 overlapping ASNs, \ours has 82,627 families vs 87,332, with 10,360 vs 8,488 multi-AS families and average size 1.36 vs 1.28.
\ours unifies 9,244 \borges families into 3,733, while \borges performs 571 reverse unifications subsuming 1,402 of \ours.
This reverse effect is driven by branding- and domain-based, partnership-heavy merges.
Because \ours requires present-tense ownership or explicit parent-child direction, it avoids cascade merges from partnership-only ties.
Manual validation shows higher precision for \ours (98.9\% vs 84.2\%) and more residuals on \borges (28) than on \ours (3).
\end{packed_itemize}

\subsubsection{Manual Validation and Error Profile}
\label{sec:manual-validation}
We now focus on unification precision; we manually validated random 100 \ours families covering 2,981 ASNs, and we sampled 50 families from each baseline families, covering 2,195 ASNs for \orgplus, 737 ASNs for \assibling, and 1,909 ASNs for \borges.

{\em The resulting TPRs are 98.9\% for \ours, 91.1\% for \orgplus, 91.9\% for \assibling, and 84.2\% for \borges.}

\subsubsection{Key takeaways}
\label{sec:key-takeaways}
\begin{packed_itemize}
	\item{Fewer families, more multi-AS groups:} Increases of 21-24\% in multi-AS families and 6-10\% in average size indicate that \ours reduces fragmentation by unifying aliases and subsidiaries split across RIR handles and sparse PeeringDB text.
		{\em We acknowledge that fewer families do not automatically imply correctness; however, manual audits and operator feedback support the quality of these unifications.}
\item{Unification factors:} When \ours subsumes a baseline, the median unification factor is about 2.5-to-1, consistent with merging small, registry-fragmented clusters into ownership-coherent families.
\item{Residuals reflect scope:} Residuals are near zero against \caida and small for \orgplus and \assibling, but larger for \borges due to its inclusion of partnership-driven ties that \ours intentionally excludes.
\end{packed_itemize}

\subsubsection{Case Studies where Web Evidence Changes the Outcome} \label{sec:comparison-cases}
We highlight three representative cases where incorporating web evidence corrected or enriched organizational groupings that previous approaches, constrained to WHOIS or PeeringDB, could not capture.

\begin{packed_itemize}
	\item{Microsoft - ZeniMax - id Software:}
\ours reconstructs a two-level hierarchy: id Software \(\rightarrow\) ZeniMax \(\rightarrow\) Microsoft'' using present-tense ownership statements, correctly mapping AS10793 under ZeniMax including its German subsidiary and then under Microsoft~\cite{microsoft-2020-zenimax, remo-2009-bethesda}.
Neither WHOIS nor PeeringDB alone consistently surfaces this cascade; the unification depends on web sources and directional validation.
\item{Orange across RIRs:} Orange's ASNs appear under multiple RIR handles such as ``Orange Business Digital Norway AS'' for AS25148, ``Orange S.A'' for AS2278, and ``Orange Business Services U.S. Inc''. for AS13879, with fragmented or absent PeeringDB entries.
\ours reconciles these into one family through aliasing and cross-RIR evidence, improving organization-level analyses.
\item{Deloitte across five RIRs and one NIR:}
Deloitte-related entities register ASNs at all five RIRs and a NIR, yet only a subset appear in PeeringDB as shown in \autoref{tbl:org-rir-case-study}.
\ours unifies these into a single family while preserving directed links to subsidiaries such as National TeleConsultants LLC, showing why registry and PeeringDB alone are insufficient.
		\end{packed_itemize}
\input{tables/org_rir_case_study}

%% file: tables/new-comparison-combined.tex
\begin{table*}[t]
\fontsize{8}{10}\selectfont
\centering
\begin{tabular}{l|c|c|c|c|c|c|c|c}
    \multicolumn{1}{c|}{\multirow{2}{*}{\textbf{Metric}}} &
    \multicolumn{2}{c|}{\textbf{\caida}} &
    \multicolumn{2}{c|}{\textbf{\orgplus}} &
    \multicolumn{2}{c|}{\textbf{\assibling}} &
    \multicolumn{2}{c}{\textbf{Borges}} \\
    & \texttt{\ours} & \caida
    & \texttt{\ours} & \texttt{AS2ORG+}
    & \texttt{\ours} & \texttt{AS Sibling}
    & \texttt{\ours} & \texttt{Borges} \\
\hline
\textbf{Common ASes} & \multicolumn{2}{c|}{111,942} & \multicolumn{2}{c|}{111,942} & \multicolumn{2}{c|}{108,902} & \multicolumn{2}{c}{111,937} \\
\textbf{Organization Families} & 82,631 & 89,486 & 82,631 & 89,280 & 82,064 & 88,129 & 82,627 & 87,332 \\
\textbf{\hspace{1em}Size $>$ 1} & 10,358 & 8,357 & 10,358 & 8,327 & 10,272 & 8,521 & 10,360 & 8,488 \\
\textbf{Avg. Family Size (\# of ASes)} & 1.36 & 1.25 & 1.36 & 1.25 & 1.33 & 1.24 & 1.36 & 1.28 \\
\textbf{Unifications by \texttt{\ours} (A / B)} & 4,272 / 11,127 & -- & 4,232 / 10,929 & -- & 3,732 / 9,824 & -- & 3,733 / 9,244 & -- \\
\textbf{Rvs. Unifications by baseline (A / B)} & -- & 0 / 0 & -- & 73 / 129 & -- & 102 / 146 & -- & 571 / 1,402 \\
\textbf{Validated Families (ASes)} & 100 (2,981) & -- & 100 (2,981) & 50 (2,195) & 100 (2,981) & 50 (737) & 100 (2,981) & 50 (1,909) \\
\textbf{True Positive Ratio} & 98.9\% & -- & 98.9\% & 91.1\% & 98.9\% & 91.9\% & 98.9\% & 84.2\% \\
\textbf{Residual Families} & 0 & 0 & 0 & 8 & 1 & 20 & 3 & 28 \\
\end{tabular}
\caption{
Comparison of \ours with four baseline datasets on overlapping AS sets.
``Unifications by \texttt{\ours}'' reports ``A / B'': \texttt{\ours} has A families that \textit{subsume} B families from the baseline.
``Reverse Unifications by baseline'' analogously reports how many baseline families subsume how many of \texttt{\ours}.
Residual families neither match nor are subsumed. 
}
\vspace{-1em}
\label{tab:compare_all}
\end{table*}

%% file: tables/org_rir_case_study.tex
\begin{table}[]
   \fontsize{8}{10}\selectfont
\begin{tabular}{ccccl}
	\textbf{ASN} & \textbf{Organization Name} & \textbf{RIR/NIR} & \textbf{Incl. PeeringDB} \\ \hline
328312    & Deloitte Touche South Africa       & AFRINIC            & Yes            \\
42536     & Deloitte LLP                       & RIPE                & No               \\
132384    & Deloitte Consulting India Pvt. Ltd & APNIC               & No                \\
55228     & National TeleConsultants LLC~\cite{beckie-2022-deloitte}    & ARIN                & No                 \\
272103    & DELOITTE TOUCHE LTDA               & LACNIC              & No            \\
131077    & Deloitte Tohmatsu Group LLC        & JPIRR               & Yes             
\end{tabular}
\caption{Deloitte's organizations appear in six different RIRs, yet only two are listed in PeeringDB.}
\vspace{-0.2in}
\label{tbl:org-rir-case-study}
\end{table}

%% file: usecase.tex
\section{Use cases and Impact}

\subsection{Organization Cone Size and Ranking}

\label{sec:org-cone-size}
Accurate estimates of an ISP's ``customer cone''
are central to many topology and ranking analyses~\cite{luckie-2013-asrelationship}.
CAIDA's existing AS Ranking dataset computes organization-level
ranks by grouping ASes in its \caida dataset and then
measuring each group's transit cone (the set of ASNs, prefixes,
or IP addresses served as customers).

Using \ours, we recreate an analogous cone-size analysis.
Among the 89,072 organizations in CAIDA's AS Rank dataset, we observe that 11,760 organizations see a larger cone size under our new mapping as \ours merges previously fragmented AS sets belonging to the same real-world operator. 
\autoref{tab:org-ranking} highlights the top 10 organizations with the largest positive changes in cone size and rank.
For example, Charter's cone size nearly doubles (from 962 to 1,819 ASes) after identifying 237 ASes under its control, raising its rank from 64 to 32.
Such improvements cannot be captured by PeeringDB-based methods alone since Charter omits details in PeeringDB's ``notes'' field, and only 8 of its 237 ASes appear in PeeringDB at all.

\subsection{RPKI Misconfiguration}
\label{sec:rpki-misconfig}

The Resource Public Key Infrastructure (RPKI) cryptographically secures
route advertisements by allowing resource holders to publish Route Origin
Authorizations (\roas). RPKI aids in preventing route hijacks, but
\emph{misconfigurations} can still arise if an organization incorrectly
assigns multiple ASNs under its umbrella. Prior work~\cite{chung-2019-rpki}
used CAIDA's \texttt{\caida} to identify RPKI-invalid prefixes announced
by the same organization's ASes.

We replicate this study using more recent data from January~2023 to July~2024,
combining BGP routing tables from RouteViews~\cite{RouteViews} and \roas
from all five RIRs, leaving 42,654 RPKI-invalid prefixes for analysis.
Applying \texttt{\caida} finds 4,436 of these as \emph{intra-organization}
misconfigurations, whereas \ours identifies an additional 1,219 such
cases (a 27.5\% increase). These newly discovered cases were previously
mis-categorized because \texttt{\caida} did not recognize that the two
ASNs belonged to the same entity.

\input{tables/org_rank.tex}

\subsection{IP Leasing Detection}
\label{sec:ip-leasing}

As IPv4 space becomes scarce, IP leasing—where organizations temporarily
rent out IP addresses—has grown into a significant market. Distinguishing
legitimate intra-organization reallocation from actual cross-organization
leasing is important for accurately characterizing these markets.

Building upon \cite{du-2024-sublet}, which used \texttt{\caida} to detect
leased addresses, we re-examined 47,318 previously flagged ``leased'' IP
prefixes. Under our new mapping, 2,783 (5.9\%) of those prefixes turn
out to be reallocated \emph{within the same} organization, rather than
leased. For example, 193.82.107.0/24 was incorrectly labeled as leased
between AS1290 and AS4637, although both are actually owned by \texttt{Telstra}.
Thus, using \ours avoids overestimating lease activity.

\subsection{Hijack Detection}
\label{sec:hijack-detection}

BGP hijacking remains a serious threat to Internet security, allowing
attackers to redirect or intercept traffic. Existing hijack detection
platforms (\eg Radar~\cite{cloudflare-radar} and GRIP~\cite{grip})
often rely on \texttt{\caida} to filter out potential false positives:
if two ASes belong to the same organization, an overlapping prefix
announcement might be legitimate rather than malicious.

We collected 17,282 hijack alerts from Radar and 11,450 from GRIP spanning
January~2023 to July~2024, none of which were flagged as intra-organization
by \texttt{\caida}. However, by applying our dataset, we find that
1,465 (8.5\%) of Radar's alarms and 1,326 (11.6\%) of GRIP's alarms
involve AS pairs owned by the same organization—thus false positives.
In total, 1,621 (9.4\%) hijack alerts across both platforms are, in fact,
benign. 
We validated 100 randomly selected cases {\em by emailing each
allegedly ``victimized'' operator's publicly listed contact}. Of the
32 who responded, \textit{all} of them confirmed the event was an internal
reannouncement rather than a hijack. This underscores how improved
AS-to-Organization mapping can substantially reduce unnecessary
alarms, saving time and effort for both operators and security teams.

%% file: tables/org_rank.tex
\begin{table}[]
    \centering
    \fontsize{8}{9}\selectfont
    \begin{tabular}{l|l|c}
\multicolumn{1}{c|}{\textbf{Org Name}} & \textbf{New Rank} &  \textbf{New Cone Size} \\
    \hline
		\textsf{TATA} & 6 (\Triangle 2) &  21,782 ($+$2,447) \\
		\textsf{Orange} & 12 ( - ) & 9,958 ($+$1,927) \\
		\textsf{Charter} & 32 (\Triangle 32) & 2,781 ($+$1,819) \\
		\textsf{Vodafone} & 14 ( - ) & 8,479 ($+$1,708) \\
		\textsf{Comcast}& 23 (\Triangle 12) & 3,862 ($+$1,513) \\
		\textsf{Liberty Global}& 21 (\Triangle 9) & 4,030 ($+$1,361) \\
		\textsf{GlobeNet Cabos}& 19 (\Triangle 8) & 4,156 ($+$1,340) \\
		\textsf{Telstra} & 13 ( - ) & 8,930 ($+$1,282) \\
		\textsf{Deutsche Telekom}& 16 (\Triangle 3) & 5,198 ($+$1,191) \\
		\textsf{Stowarzyszenie}& 44 (\Triangle 51) & 1,652 ($+$1,014) \\
    \end{tabular}
    \caption{Organization Rankings and Cone Metrics.}
	\vspace{-0.2in}
    \label{tab:org-ranking}
\end{table}

%% file: conclusion.tex
\section{Concluding Discussion} \label{sec:conclusion}
We have presented \ours, which takes a pragmatic step: bring curated web evidence into the pipeline and apply modest, well-studied NLP and LLM techniques with strict, evidence-cited validation.
The result is broader coverage and richer structure---including aliases, acquisitions, and parent-child ties---without proposing a new algorithm.
Our key findings include:

\begin{packed_itemize}
    \item \textit{Coverage and Aggregation:} Analyzing over 112,000 ASNs, 
    \ours produces 82,840 organization families, merging thousands of 
    AS2Org entries into fewer, larger families and revealing additional 
    intra-organizational relationships. 

    \item \textit{Operational Impact:} By more accurately grouping ASNs, 
    \ours increases the detection rate of intra-organizational RPKI 
    misconfigurations by 27.5\% and reduces false alarms in hijack 
    detection by 9.4\%. It also decreases misclassified IP leasing 
    cases by 5.9\%.

	\item \textit{Sustained accuracy via operator-in-the-loop.} Registries are incomplete and organizations evolve; cross-country name collisions make errors likely without external confirmation. We run a public dashboard and APIs where network operators review per-ASN evidence and submit lightweight feedback.

\end{packed_itemize}

%% file: ethics.tex
\section{Ethical Considerations} 
\label{sec:ethics}
\ours collects raw data from Internet, involves the use of web searching and crawling as the underlying data collection technique. We acknowledge that web crawling can raise ethical concerns, particularly regarding the potential for overloading servers or violating terms of service. To mitigate these risks, we follow the Menlo Report principles~\cite{dittrich-2012-menlo}, and implement several best practices in our data collection process:

\begin{packed_itemize}
    \item Respecting Robots.txt: We respect the \texttt{robots.txt} files of the websites we crawl, which specify the rules for web crawlers regarding which parts of the site can be accessed.
    \item Rate Limiting: We have implemented rate limiting in our crawling process to avoid overwhelming any single server with requests. This helps to ensure that our crawling activities do not disrupt the normal operation of the websites we access.
    \item User-Agent Identification: We identify our crawler with a specific user-agent string, allowing website administrators to recognize our crawler and take action if necessary.
    \item Data Usage: The data collected is used solely for research purposes and is not shared with third parties without proper anonymization or aggregation.
\end{packed_itemize}

%% file: reproduce.tex
\section{REPLICABILITY}
\label{sec:replicability}
We public \ours code and analysis scripts at
\begin{center}
    \url{https://asint.netsecurelab.org}
\end{center}

for network operators, administrators, and researchers to reproduce our work. 

We also provide a public available dataset for \ours AS-to-Organization mapping dataset, continuously updated with new information.

%% file: prompt.tex
\section{Prompt} 
\label{sec:prompt}

\autoref{fig:prompt1} illustrates the prompts used in organization link discovery (\autoref{sec:candidate-links}), while \autoref{fig:prompt4} presents an example prompt from organization family refinement (\autoref{sec:org_family_refinement}). Only partial and simplified versions of the prompts are shown due to space limitations; the complete prompt set will be made available at
\begin{center}
  \url{https://asint.netsecurelab.org}
\end{center}

\begin{figure*}[t]
\centering
\begin{tcolorbox}[
  title=Few-Shot Prompt,
  fonttitle=\bfseries,
  colback=gray!5!white,
  colframe=gray!75!black,
  width=\textwidth,
  boxrule=0.5pt,
  fontupper=\normalsize\rmfamily   
]

\small
\begin{verbatim}
You are an expert at determining how organizations that own or control Autonomous System (AS) numbers in computer 
networks are related, using the provided context.

You will receive:
1) A **base_organization**.
2) A list of **candidate_organizations**.
3) **context** providing relevant organizational details.

### Definitions
For each (base_organization, candidate_organization) pair, decide which of the following relationships 
best applies:

- Alias
  Both names refer to exactly the same legal entity or one is a historical name of the other.

- Parent/Subsidiary
  One organization has acquired or holds more than 50% of the stock of the other. Identify which one is the parent:
  Choose between "base_organization" or "candidate_organization".

- No_relation
  There is insufficient evidence of alias, ownership or acquisition linking them.

**Mandatory JSON Output Format**:

/******** Omitted due to paper page limits. ********/

### Example:
Provide this JSON object for each (base_organization, candidate_organization) pair as an array for output. 
Examples:
[
   {
      "base_org_name": "Zayo Bandwidth",
      "candidate_org_name": "company",
      "reasoning for Alias": "The candidate name is generic and lacks direct evidence connecting it 
      						  to Zayo Bandwidth.",
      "reasoning for Parent/Subsidiary": "No indication of ownership or acquisition.",
      "relationship": "No_relation",
      "parent": "",
      "parent name": ""
   }
]  

/******** More examples omitted due to paper page limits. ********/
 
Now, respond by considering each candidate_organization in the list, applying reasoning, and returning your 
final JSON array with one object per candidate.
\end{verbatim}
\end{tcolorbox}
\caption{Full prompt used for few-shot inference.}
\label{fig:prompt1}
\end{figure*}

\begin{figure*}[t]
\centering
\begin{tcolorbox}[
  title=Child/Parent Org Validation,
  fonttitle=\bfseries,
  colback=gray!5!white,
  colframe=gray!75!black,
  width=\textwidth,
  boxrule=0.5pt,
  fontupper=\normalsize\rmfamily  
]

\small
\begin{verbatim}
### Task
You are given context about two organizations. Your job is to determine whether there is a **current 
parent-child relationship** between them — where one **currently owns, controls, or oversees** the other.

### You will receive:
- **Organization A**
- **Organization B**
- **Context**: background information (such as web content) about either or both organizations

### Your job is to assess one of these **directional** relationships:
1. **Organization A is the parent of Organization B**  
   (A currently owns, controls, or oversees B)  
2. **Organization B is the parent of Organization A**  
   (B currently owns, controls, or oversees A)  
3. **No parent-child relationship exists**

### Strict Evaluation Criteria
You **must not infer or assume anything** beyond the context. Follow these rules:
1. Use only current relationship indicators
- Ignore past-tense relationships (e.g., “was acquired by”, “used to be a subsidiary of”) unless 
  the **present-tense relationship is confirmed elsewhere**.
- Accept only statements that **clearly describe a current relationship** (e.g., “is a subsidiary of”, 
  “is owned by”, “is part of”).
- If one organization was acquired in the past, but is no longer owned, treat it as no parent-child relationship.

2. Require exact organization name matches
- Do **not** infer connections based on partial matches or name similarity.
- Only accept relationships where the **exact organization names** (as given) appear in the relationship sentence.

3. Directionality must be explicit
- You must clearly identify which organization is the **parent** and which is the **child**, 
based on direct statements.
- Reverse relationships (e.g., “B owns A” vs. “A owns B”) must result in different scores.

### Output Format

/******** Omitted due to paper page limits. ********/

### Self-Check Before Submitting:
- Are you using only **present-tense** relationship cues?  
- Are you using **exact org name matches only**?  
- Is the **directionality** (parent vs. child) reflected correctly in the score fields?  
- If no valid match, is `No_Relationship_Score = 1.0`?  
- If a government org is involved, are all scores set appropriately?

### Example

/******** More examples omitted due to paper page limits. ********/

### input
- **Organization A**: {org_a}
- **Organization B**: {org_b}
- **Context**: {context}

### Now start your reasoning
\end{verbatim}
\end{tcolorbox}
\caption{Full prompt used for child/parent org validation}
\label{fig:prompt4}
\end{figure*}

%% file: algorithm.tex
\section{Calculation of Precision, Recall, Specificity, NPV and Accuracy}
\label{sec:calculation}

\[
\begin{aligned}
&\text{Given: } TP = 1{,}054,\; FP = 43,\; FN = 9,\; TN = 991. \\[3pt]
&\text{Recall} = \frac{TP}{TP + FN}
= \frac{1{,}054}{1{,}063}
= 0.9915 \; (99.2\%), \\[4pt]
&\text{Specificity} = \frac{TN}{TN + FP}
= \frac{991}{1{,}034}
= 0.9584 \; (95.8\%), \\[4pt]
&\text{Precision} = \frac{TP}{TP + FP}
= \frac{1{,}054}{1{,}097}
= 0.9608 \; (96.1\%), \\[4pt]
&\text{NPV} = \frac{TN}{TN + FN}
= \frac{991}{1{,}000}
= 0.9910 \; (99.1\%), \\[4pt]
&\text{Accuracy} = \frac{TP + TN}{TP + TN + FP + FN}
= \frac{2{,}045}{2{,}097}
= 0.9752 \; (97.5\%).
\end{aligned}
\]

%% file: examples.tex
\section{Examples of Wrong Inference in Candidate Organization Link Discovery (\autoref{sec:candidate-links})}
\label{sec:false_cluster}

Below is a typical example of a wrong cluster caused by name-based clustering:

\begin{packed_itemize}
  \item \textbf{Org 1: ZeniMax}
  \begin{packed_itemize}
    \item Context: \textbf{Bethesda Softworks} was reorganized as a division of \textbf{ZeniMax}. 
          By then Bethesda employed nearly 100 people.
    \item NER output: [Bethesda, ZeniMax]
    \item LLM: Bethesda is child.
  \end{packed_itemize}

  \item \textbf{Org 2: Bethesda Wireless and Fiber}
  \begin{packed_itemize}
    \item Context: \textbf{Bethesda Wireless and Fiber} offers a truly unlimited broadband internet connection.
    \item NER output: [Bethesda]
    \item LLM: Bethesda is an alias.
  \end{packed_itemize}
\end{packed_itemize}

Although the context clearly states that Bethesda Softworks is a child of ZeniMax, name-based clustering incorrectly links Bethesda Wireless and Fiber with ZeniMax as a child–parent pair. 
The error arises from incomplete NER extraction, which gives the LLM only a partial view of organization names and leads to misinference that cannot be resolved by prompt engineering alone. 
Lenient inference criteria preserve many true links but also admit false clusters, while stricter criteria reduce false positives at the cost of discarding true links. In our implementation, we adopt a lenient setting and defer removal of false clusters in later stage \autoref{sec:org_family_refinement}.

%% file: paper.bbl
\newcommand{\etalchar}[1]{$^{#1}$}
\begin{thebibliography}{00}
\bibitem{faiss}
Faiss.
\newblock \url{https://github.com/facebookresearch/faiss}.

\bibitem{grip}
GRIP - Global Routing Intelligence Platform.
\newblock \url{https://grip.inetintel.cc.gatech.edu}.

\bibitem{all-MiniLM}
all-MiniLM-L6-v2.
\newblock \url{https://huggingface.co/sentence-transformers/all-MiniLM-L6-v2}.

\bibitem{arturi-2023-as2org}
A. Arturi, E. Carisimo, and F.~E. Bustamante.
\newblock as2org+: Enriching as-to-Organization Mappings with PeeringDB.
\newblock \emph{PAM}, 2023.

\bibitem{beckie-2022-deloitte}
B. BARRETT.
\newblock Deloitte Consulting acquires National TeleConsultants in media enterprise push.
\newblock 2022.
\newblock \url{https://erp.today/deloitte-consulting-acquires-national-teleconsultants-in-media-push/}.

\bibitem{rfc6810}
R. Bush and R. Austein.
\newblock The Resource Public Key Infrastructure (RPKI) to Router Protocol.
\newblock RFC 6810, IETF, 2013.

\bibitem{chen-2024-unleashing}
B. Chen, Z. Zhang, N. Langrené, and S. Zhu.
\newblock Unleashing the potential of prompt engineering in Large Language Models: a comprehensive review.
\newblock 2024.
\newblock \url{https://arxiv.org/abs/2310.14735}.

\bibitem{chung-2019-rpki}
T. Chung, E. Aben, T. Bruijnzeels, B. Chandrasekaran, D. Choffnes, D. Levin, B.~M. Maggs, A. Mislove, R. van Rijswijk-Deij, J.~P. Rula, and N. Sullivan.
\newblock RPKI is Coming of Age: A Longitudinal Study of RPKI Deployment and Invalid Route Origins.
\newblock \emph{IMC}, 2019.

\bibitem{cai-2010-as}
X. Cai, J.~S. Heidemann, B. Krishnamurthy, and W. Willinger.
\newblock Towards an as-to-Organization Map.
\newblock \emph{IMC}, 2010.

\bibitem{chen-2023-sibling}
Z. Chen, Z.~S. Bischof, C. Testart, and A. Dainotti.
\newblock Improving the Inference of Sibling Autonomous Systems.
\newblock \emph{PAM}, 2023.

\bibitem{CaidaASOrganization}
CAIDA {AS}Organizations Dataset.
\newblock \url{http://www.caida.org/data/as-organizations/}.

\bibitem{cloudflare-radar}
Cloudflare Radar.
\newblock \url{https://radar.cloudflare.com/routing}.

\bibitem{du-2024-sublet}
B. Du, R. Fontugne, C. Testart, A.~C. Snoeren, and k. claffy.
\newblock Sublet Your Subnet: Inferring IP Leasing in the Wild.
\newblock \emph{IMC}, 2024.

\bibitem{dittrich-2012-menlo}
D. Dittrich and E. Kenneally.
\newblock The Menlo Report: Ethical Principles Guiding Information and Communication Technology Research.
\newblock 2012.
\newblock \url{https://www.dhs.gov/sites/default/files/publications/CSD-MenloPrinciplesCORE-20120803_1.pdf}.

\bibitem{gao-2023-retrieval}
Y. Gao, Y. Xiong, X. Gao, K. Jia, J. Pan, Y. Bi, Y. Dai, J. Sun, M. Wang, and H. Wang.
\newblock Retrieval-Augmented Generation for Large Language Models: A Survey.
\newblock 2023.
\newblock \url{https://arxiv.org/abs/2312.10997}.

\bibitem{husain-2019-codesearchnet}
H. Husain, H.-H. Wu, T. Gazit, M. Allamanis, and M. Brockschmidt.
\newblock CodeSearchNet Challenge: Evaluating the State of Semantic Code Search.
\newblock 2019.
\newblock \url{https://arxiv.org/abs/1909.09436}.

\bibitem{huang-2025-hallucination}
L. Huang, W. Yu, W. Ma, W. Zhong, Z. Feng, H. Wang, Q. Chen, W. Peng, X. Feng, B. Qin, and T. Liu.
\newblock A Survey on Hallucination in Large Language Models: Principles, Taxonomy, Challenges, and Open Questions.
\newblock \emph{ACM TOIS}, 43, 2025.

\bibitem{jing-2022-ner}
L. Jing, S. Aixin, H. Jianglei, and L. Chenliang.
\newblock A Survey on Deep Learning for Named Entity Recognition.
\newblock \emph{IEEE Transactions on Knowledge and Data Engineering}, 34(1), 2022.

\bibitem{kaddour-2023-challenges}
J. Kaddour, J. Harris, M. Mozes, H. Bradley, R. Raileanu, and R. McHardy.
\newblock Challenges and Applications of Large Language Models.
\newblock 2023.
\newblock \url{https://arxiv.org/abs/2307.10169}.

\bibitem{bert-base-NER}
D. Lim.
\newblock Bert-base-NER.
\newblock \url{https://huggingface.co/dslim/bert-base-NER}.

\bibitem{luckie-2013-asrelationship}
M. Luckie, B. Huffaker, K. Claffy, A. Dhamdhere, and V. Giotsas.
\newblock AS Relationships, Customer Cones, and Validation.
\newblock \emph{IMC}, 2013.

\bibitem{lewis-2020-retrieval}
P. Lewis, E. Perez, A. Piktus, F. Petroni, V. Karpukhin, N. Goyal, H. Küttler, M. Lewis, W.-t. Yih, T. Rocktäschel, S. Riedel, and D. Kiela.
\newblock Retrieval-Augmented Generation for Knowledge-Intensive NLP Tasks.
\newblock \emph{NIPS}, 2020.

\bibitem{li-2023-rag}
X. Li, Z. Liu, C. Xiong, S. Yu, Y. Gu, Z. Liu, and G. Yu.
\newblock Structure-Aware Language Model Pretraining Improves Dense Retrieval on Structured Data.
\newblock \emph{ACL}, 2023.

\bibitem{microsoft-2020-zenimax}
Microsoft.
\newblock Microsoft finalizes acquisition of ZeniMax Media.
\newblock 2020.
\newblock \url{https://news.microsoft.com/features/microsoft-finalizes-acquisition-of-zenimax-media/}.

\bibitem{PeeringDB}
PeeringDB: The Interconnection Database.
\newblock \url{https://www.peeringdb.com/}.

\bibitem{remo-2009-bethesda}
C. Remo.
\newblock Bethesda Parent ZeniMax Acquires id Software.
\newblock 2009.
\newblock \url{https://www.gamedeveloper.com/game-platforms/bethesda-parent-zenimax-acquires-id-software}.

\bibitem{RouteViews}
University of Oregon RouteViews project.
\newblock \url{http://www.routeviews.org/}.

\bibitem{selmo-2025-borge}
C. Selmo, E. Carisimo, F.~E. Bustamante, and J.~I. Alvarez-Hamelin.
\newblock Learning as-to-Organization Mappings with Borges.
\newblock \emph{IMC}, 2025.

\bibitem{shuster-2021-retrieval}
K. Shuster, S. Poff, M. Chen, D. Kiela, and J. Weston.
\newblock Retrieval Augmentation Reduces Hallucination in Conversation.
\newblock 2021.
\newblock \url{https://arxiv.org/pdf/2104.07567}.

\bibitem{sahoo-2024-systematic}
P. Sahoo, A.~K. Singh, S. Saha, V. Jain, S. Mondal, and A. Chadha.
\newblock A Systematic Survey of Prompt Engineering in Large Language Models: Techniques and Applications.
\newblock 2024.
\newblock \url{https://rotmandigital.ca/wp-content/uploads/2024/09/A-Systematic-Survey-of-Prompt-Engineering-in-Large-Language-Models.pdf}.

\bibitem{Milvus}
The High-Performance Vector Database Built for Scale.
\newblock \url{https://milvus.io/}.

\bibitem{xu-2024-rag}
P. Xu, W. Ping, X. Wu, L. McAfee, C. Zhu, Z. Liu, S. Subramanian, E. Bakhturina, M. Shoeybi, and B. Catanzaro.
\newblock Retrieval meets Long Context Large Language Models.
\newblock \emph{iclr}, 2024.

\bibitem{rfc4271}
R. Yakov, L. Tony, H. Susan, and others.
\newblock A border gateway protocol 4 (BGP-4).
\newblock RFC 4271, IETF, 1994.

\bibitem{zhao-2021-calibrate}
Z. Zhao, E. Wallace, S. Feng, D. Klein, and S. Singh.
\newblock Calibrate Before Use: Improving Few-shot Performance of Language Models.
\newblock \emph{ICML}, 2021.

\end{thebibliography}
